\documentclass[useAMS,usenatbib]{mn2e}

\usepackage{graphicx, natbib,epsfig,amsmath,amssymb, mathrsfs, txfonts} %
\DeclareGraphicsExtensions{.eps, .jpg, .ps}

\usepackage{epstopdf}

\title[Seeing estimation from Shack-Hartmann images]
  {Active optics Shack-Hartmann sensor: using spot sizes to measure the seeing at the focal plane of a telescope}
\author[P. Martinez et al.]
  {P.~Martinez,$^{1}$ \thanks{patrice.martinez@obs.ujf-grenoble.fr}
  J.~Kolb,$^2$ M.~Sarazin,$^2$ J.~Navarrete $^3$
 \\
  $^1$ UJF-Grenoble 1 / CNRS-INSU, Institut de Plan\'{e}tologie et d'Astrophysique de Grenoble (IPAG) UMR 5274, Grenoble, F-38041, France \\  
  $^2$ European Southern Observatory, Karl-Schwarzschild-Strasse 2, D-85748, Garching, Germany \\
  $^3$ European Southern Observatory, Alonso de Cordova 3107, Casilla 19001, Vitacura, Santiago, Chile\\}
\date{Released 2011 Xxxxx XX}

\pagerange{\pageref{firstpage}--\pageref{lastpage}} \pubyear{2011}

\def\LaTeX{L\kern-.36em\raise.3ex\hbox{a}\kern-.15em
    T\kern-.1667em\lower.7ex\hbox{E}\kern-.125emX}

\begin{document}

\label{firstpage}

\maketitle

\begin{abstract}
Real-time seeing estimation at the focus of a telescope is nowadays strongly emphasized as this knowledge virtually drives the dimensioning of adaptive optics systems and instrument operational aspects. In this context we study the interest of using active optics Shack-Hartmann (AOSH) sensor images to provide accurate estimate of the seeing.  The AOSH practically delivers long exposure spot PSFs -- at the critical location of the telescope focus -- being directly related to the atmospheric seeing in the line of sight.
Although AOSH sensors are not specified to measure spot sizes but slopes, we show that accurate seeing estimation from AOSH images can be obtained with a dedicated algorithm. The sensitivity and comparison of two algorithms to various parameters is analyzed in a systematic way, demonstrating that efficient estimation of the seeing can be obtained by adequate means. 
\end{abstract}

\begin{keywords}
Site testing  -- atmospheric effects -- methods: numerical -- methods: data analysis
\end{keywords}

\section{Introduction}
The evaluation of the seeing is paramount for selecting astronomical sites and following their temporal evolution.
Likewise, its estimation is fundamental for the dimensioning of adaptive optics systems and their performance predications.
This knowledge virtually drives instruments operational aspects at a telescope, and more emphasis is made to develop and use accurate real-time seeing estimator at the focus of a telescope.

The atmospheric seeing is commonly measured by the differential image motion monitor \citep[DIMM,][]{Sarazin90} in most observatories, or by means of alternative seeing monitors, e.g., Generalized Seeing Monitor \citep[GSM,][]{Ziad00}, Multi-Aperture Scintillation Sensor \citep[MASS,][]{Kornilov01}.
Being localized away from the telescope platform, such a device delivers seeing estimate that can significantly differ from the \textit{effective} seeing seen at a telescope focii because of  pointing orientation and/or height above the ground differences, or local seeing bias (dome contribution). The effect of the two latter will largely expand in the context of the next generation of telescopes: the extremely large telescopes.  

On the other hand, the atmospheric seeing can be deduced at the focus of a large telescope from the width of the long-exposure point spread function (PSF) assuming successive corrections to apply (turbulence outer scale, wavelength, and airmass). We remind the reader that the seeing is defined at $\lambda$ = 500nm for observations at zenith.
For this purpose, several flavors of images can be used at the critical location of the telescope focii: (1/) scientific instrument images, (2/) guide probe images, (3/) active optics Shack-Hartmann images.
At the VLT,  focal planes are equipped with an arm used for acquisition of a natural guide star. The light from this star is then split between a guide probe for an accurate tracking of the sky, and a Shack-Hartmann wavefront sensor used by the active optics to control the shape of the primary mirror.

However seeing estimation from the full-width at  half  maximum (FWHM) of a PSF strongly relies on the exposure time that must be long enough so that the turbulence has been averaged; ensuring that all representations of the wavefront spatial scales have passed through the pupil. This is thus dependent on pupil size and turbulence velocity, though it is commonly admitted that 30 seconds average properly the turbulence, introducing significant FWHM biases otherwise.
This already discards the use of the guide probe as exposure times are not longer than 50 milli-second, while scientific instrument images do not entirely comply with the real-time aspect of the seeing estimation. Instruments are affected by an observational bias:  unavailability for a large range of seeing conditions, and are additionally affected by the telescope field stabilization. 

Active Optics Shack-Hartmann (AOSH) delivers continuously real-time images of long exposure spot PSFs (typically 45 seconds) at the same location of scientific instruments.
AOSH images provide simultaneously various data: slopes, intensities, and spot sizes. When short exposures are used, the information provided by both slopes and intensities (i.e., scintillation) can be used to retrieve $Cn^{2}$ profile using correlations of these data from two separated stars \citep{Robert11}. In this paper, we propose to use the spot sizes in the sub-apertures to retrieve the atmospheric seeing in the line of sight. 
In other words, we propose to use the active optics AOSH sensor system of the telescope as a turbulence monitor to provide accurate seeing estimation directly at the telescope focus.
For this purpose, careful assumption of the long-exposure PSF profile (to ensure precise FWHM evaluation), and accurate derivation of the seeing from the estimated FWHM are required. 

In this context, we investigate the performance and limitations of two different methods to estimate the seeing from long-exposure AOSH spot PSFs. 
The comparison and selection of the uppermost modus operandi is a byproduct of this work. 
Our study is carried out by means of extensive simulations, and using real data from the AOSH images obtained at the VLT. 
In Sect. 2 we describe two different algorithms developed to retrieve the seeing from long exposure images, and in Sect. 3  we present our simulation hypothesis, and discuss our results. In Sect. 4 real-data from the VLT AOSH database are re-analyzed and compared to synchronous image measurements obtained at the VLT (UT4) with the instrument FORS2 \citep{FORS2}. 
Finally in Sect. 5 we draw conclusions.

\section{Extracting seeing from Shack-Hartmann spot sizes}
Accurate seeing estimation from the full-width at  half  maximum (FWHM) of a turbulence-limited long exposure PSF requires two conditions: (1/) careful assumption of the long-exposure PSF profile to ensure precise FWHM evaluation; (2/) accurate derivation of the seeing from the estimated FWHM. In the following, we treat these two aspects, and present in details the two dedicated algorithms.

\subsection{Long-exposure PSF profile}
The theoretical expression of a long-exposure PSF can be described through the expression of its optical transfer function (OTF) obtained by multiplying the
telescope OTF, denoted $T_{0}(\bf{f})$, by the atmospheric OTF:
\begin{equation}
T_a ({\bf f}) =  \exp [ -0.5 D_{\phi}(\lambda {\bf f})], 
\label{eq:Tf}
\end{equation}
where  ${\bf  f}$ is the angular spatial frequency, $\lambda$ is the  imaging wavelength, and $ D_{\phi}({\bf  r})$ is the phase structure  function \citep{Good85, Roddier81}.  This expression is miscellaneous and applicable to any turbulence spectrum and any telescope diameter. \\
\noindent The analytic  expression for the  phase structure function in the Kolmogorov-Obukhov model can be found in \citet{T61} and is expressed by $ D_{\phi}(r) = 6.88 (r/r_0)^{5/3}$, where $r_0$ is the Fried's coherence radius \citep{Fried66}.  
Finally, the long-exposure OTF can be expressed as:
\begin{equation}
T(\textbf{f}) = T_{0} (\textbf{f}) \times exp[-3.44 (\lambda \textbf{f}/r_{0})^{5/3}],
\label{A1} 
\end{equation}
and long-exposure PSFs are accurately described by Eq.~\ref{A1}.
In the case of a  large ideal telescope with diameter $D \gg
r_0$ the  diffraction term $T_{0}$ can be neglected, while in the case of AOSH sub-apertures of size $d$ it cannot ($d\approx r_0$).  

We note that Eq.~\ref{A1} assumes nonrealistic behavior of the low-frequency of the turbulence model phase spectrum. 
It is firmly established that the phase spectrum deviates from the power law at low frequencies \citep{Ziad00, Toko07}, and 
this behavior is described in a first order by an additional parameter, the outer scale $L_0$. The expression for the  phase structure function ($D_{\phi}$) with finite outer scale $L_0$ can be found in \citep[e.g.,][]{T61, Toko2002}. As large-scale wavefronts are anything but stationary, AOSH long-exposure spot PSFs can be in a first order described by Eq.~\ref{A1}, while a posteriori correction by $L_0$ experimental estimate is mandatory. The examination of the $L_0$ influence will be treated in the following.

\begin{figure}
\centering
\includegraphics[width=6.0cm]{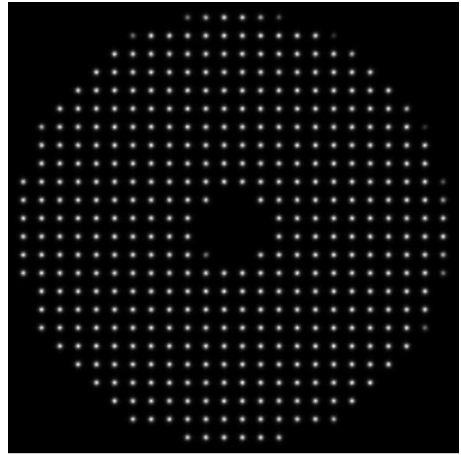}
\caption{Simulated AOSH image based on the VLT AOSH geometry.}
\label{model}
\end{figure}

\subsection{Seeing and FWHM}
The Kolmogorov-Obukhov model predicts dependence of the PSF FWHM  $\varepsilon_{0}$  on  wavelength
$\lambda$ and $r_0$:
\begin{equation}
\varepsilon_{0} = 0.976 \; \lambda /r_{0} .
\label{eq1}
\end{equation}
Equation \ref{eq1} is the definition of the seeing (assumed at $\lambda = 500$\,nm, and for observations at zenith).
However, the  physics of  turbulence implies  that the  spatial  power spectral density (PSD) of phase distortions  $W_{\phi} ({\bf v})$ deviates from the pure power law at low  frequencies.  A popular  von  K\`arm\`an  (vK) turbulence  model \citep[e.g.,][]{T61, Ziad00} introduces an additional parameter, the {\em outer scale} $L_0$ that describes the low-frequency behavior of the turbulence model phase spectrum:
\begin{equation}
W_{\phi}  ({\bf v}) = 0.0229\;  r_0^{-5/3} \; ( |{\bf v}|^2 + L_0^{-2}  )^{-11/6}.
\label{eq:L0}
\end{equation}
Equation~\ref{eq:L0} is the definition  of $L_0$, where ${\bf v}$ is the spatial frequency in m$^{-1}$. 
The Kolmogorov-Obukhov model corresponds to  $L_0 = \infty$.  In  the vK model,  $r_0$ describes the high-frequency asymptotic behavior of the spectrum.
In this context, the outer scale of the turbulence ($L_0$) plays a significant role in the improvement of image quality (i.e., FWHM) at the focus of a telescope. The image quality is  different (and in some cases by a large factor, e.g., 30$\%$ to 40$\%$ in the near-infrared) from the atmospheric seeing that can be measured by dedicated seeing monitors, such as the DIMM.\\
The dependence of atmospheric long exposure resolution on $L_0$ is efficiently predicted by a simple approximate formula (Eq. \ref{toko}) introduced by \citet{Toko2002}, and confirmed by means of extensive simulations \citep{2010A&A...516A..90M, 2010Msngr.141....5M}, where we emphasized that the effect of  finite $L_0$ is independent of the telescope diameter. The validity of Eq. \ref{toko} has been established in a $L_0$/$r_0$$>$20 and $L_0/D$$\leq$500 domain, where our treatment of the diffraction failed for small telescope diameters \citep[D$<$1m, ][]{2010A&A...516A..90M}. 
\begin{equation}
FWHM \approx \varepsilon_{0} \; \sqrt{1 - 2.183 ( r_{0} / L_{0})^{0.356}},
\label{toko}
\end{equation}
As a consequence, to deduce atmospheric seeing $\varepsilon_{0}$ (at 500nm) from the FWHM of a long-exposure PSF the correction implied by Eq. \ref{toko} is mandatory prior to airmass and wavelength correction. 
We will show that this does concern even the case of small size ($d$) AOSH sub-apertures, where $L_{0}$ $\gg$ $d$. 

\subsection{Algorithms principle}
We consider two different algorithms to extract the FWHM of a long exposure AOSH image. These algorithms differ from their assumptions on the sub-aperture spot profile and the way sub-aperture diffraction is accounted for, while they rely on a common preliminary step for the selection of spots retained for the analysis. 

\subsubsection{AOSH spots selection}
For both algorithms, background estimation is performed on a corner of the image without spots, and hot pixels are set to the background.
The cleanest and un-vignetted spots are selected in each frame for the analysis. These extracted spots are oversampled by a factor two, re-centered and averaged. The averaging reduces the influence of potential local CCD defects (e.g, bad pixels, bias structures, etc...). In practice, the averaged spot is based on hundreds of selected spots.

\subsubsection{OTF-based algorithm}
The first algorithm, hereafter A1, has been proposed by \citet{Toko07} and is based on the long-exposure spot PSF profile defined in Eq. \ref{A1}.  The modulus of the long-exposure optical transfer function of the averaged spot is calculated and normalized. It is then divided by the square sub-aperture diffraction-limited transfer function $T_0(\bf{f}$): 
\begin{equation}
T_{0}(\bf{f}) = (1-[\lambda \bf{f_x}/d]) \times (1 - [\lambda \bf{f_y}/d])
\label{diff1}
\end{equation}
where $d$ is the size of the AOSH sub-aperture. 
At this stage the cut of the $T(\bf{f})$ along each axis can be extracted and fitted to the exponential part of Eq. \ref{A1} to derive a single parameter $r_0$, or equivalently Fourier transformed to derive the FWHM of the resulting spot PSF profile using a 2-dimensional elliptical Gaussian fit. The orientation of the long and small axis of $T(\bf{f})$ is found by fitting it with a 2-dimensional elliptical Gaussian.

\subsubsection{PSF-based algorithm}
The second algorithm, hereafter A2, has been proposed by \citet{Lothar06}, and is in use as a diagnostic tool at the VLT since May 2010.
A2 relies on the assumption that AOSH spots can be described by the following rotationally symmetric PSF profile:
\begin{equation}
F(\textbf{u}) \propto F_{0}(\textbf{u}) \otimes exp[-(\textbf{u}/r_{0})^{5/3}]
\label{A2}
\end{equation}
where $\bf{u}$ is the spatial coordinate, $F_{0}$ the diffraction limited PSF of the AOSH sub-aperture. The symbol $\otimes$ denotes the convolution product. 
A2 relies on an approximate PSF profile based on the analytical expression of the OTF defined in  Eq.Ê\ref{A1}.
The algorithm A2 works as follow: an initial estimation of the FWHM (hereafter, $\theta$) is derived by fitting the averaged spot profile to the exponential part of Eq.Ê\ref{A2}.
Then A2 approximately accounts for $F_0(\bf{u})$ by quadratically subtracting the sub-aperture diffraction $\theta_{0} = \lambda/d$ from the estimated FWHM:
\begin{equation}
FWHM \approx \sqrt{{\theta}^{2} - {\theta_{0}}^{2}} .
\label{diff2}
\end{equation}
This quadratic subtraction defined in Eq. \ref{diff2} is justified by the assumption that the two terms of Eq. \ref{A2} can be approximated to Gaussian functions. 

\begin{figure}
\includegraphics[width=8.2cm]{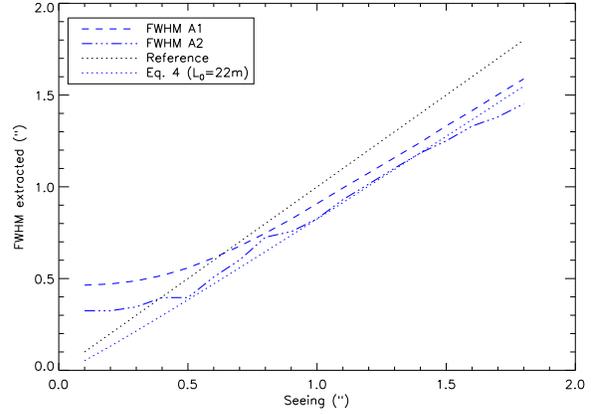}
\caption{FWHM estimation from simulated AOSH images as function of the seeing obtained with A1 and A2.}
\label{comp}
\end{figure}

\section{Calibration and results}
In order to calibrate and compare these two algorithms, we test them with simulated AOSH images. To match our simulations to real situation, we assume identical conditions as the ones encountered with the AOSH system of the VLT. 
Assumptions for the AOSH model and atmospheric turbulence are described bellow. The wavelength considered through the study is 500nm except when the effect of the wavelength is analyzed.

\subsection{Simulation hypothesis}

\subsubsection{Shack-Hartmann model}
Our simulations are based on a diffractive Shack-Hartmann model that reproduces the VLT AOSH geometry: 24 sub-apertures across the pupil diameter, 22 pixels per sub-aperture, 0.305$\arcsec$ pixel scale, $d = D/24 = 0.338$ m. The validity of the AOSH model has been verified through several aspects such as the plate scale, spot sizes, slope measurements, and phase reconstruction. 

\subsubsection{Atmospheric turbulence}
The atmospheric turbulence is simulated with 300 uncorrelated phase screens of dimension $3072 \times 3072$ pixels (i.e. 45 meters width). 
The principle of the generation of a phase screen is based on the Fourier approach: randomized white noise maps are colored in the Fourier space by the turbulence power spectral density (PSD) function, and the inverse Fourier transform of an outcome correspond to a phase screen realization.
 
The validity of the atmospheric turbulence statistic has been carried out on the simulated phase screens:  the value of the outer scale ($L_{0}$), Fried parameter ($r_{0}$) and seeing of the phase screens have been confirmed by decomposition on the Zernike polynomials and variance measurements over the 300 uncorrelated phase screens. In addition the validity of the long-exposure AOSH image has been verified. In Fig. \ref{model}, we show an example of a simulated AOSH image. 

We have generated AOSH images through atmospheric turbulence with seeing conditions from 0.1 to 1.8$\arcsec$ with 0.1$\arcsec$ increment. Except when the turbulence outer scale is an open-parameter, $L_0$ is always defined at 22 meters -- VLT Paranal median value, see for instance \citet[][]{2010A&A...524A..73D}. 
\begin{figure*}
\includegraphics[width=8.2cm]{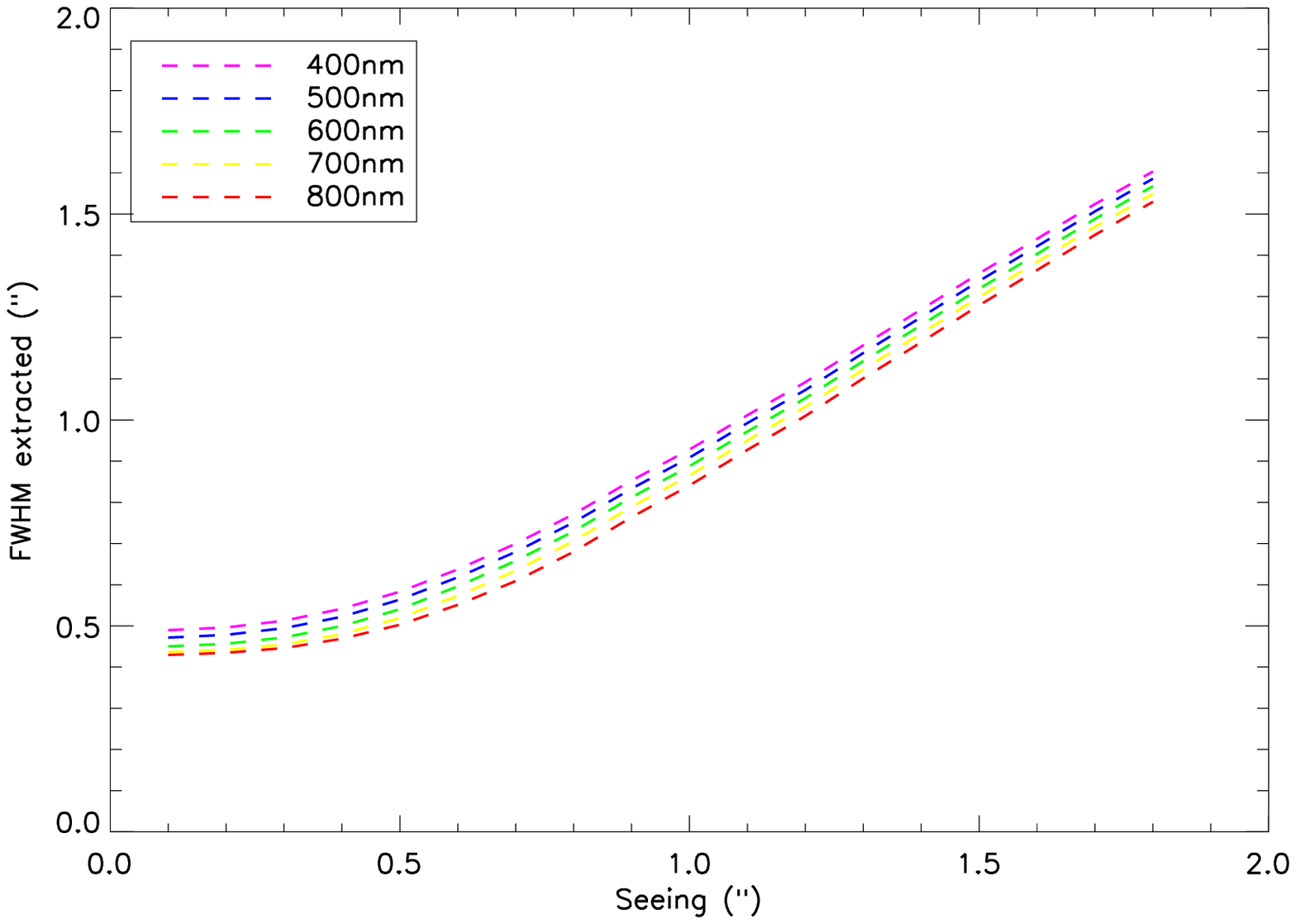}
\includegraphics[width=8.2cm]{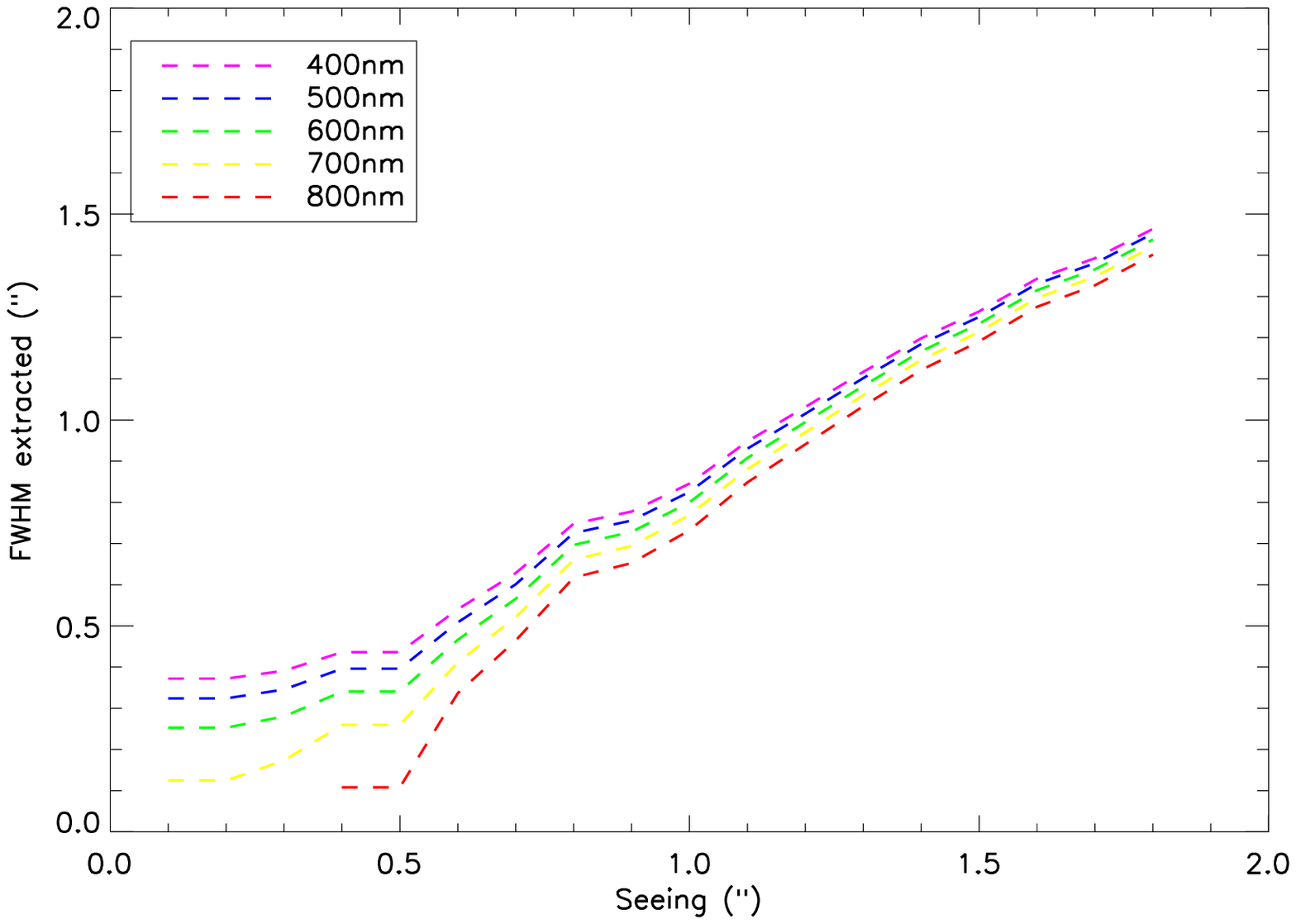}
\caption{Sub-aperture diffraction removal as function of the imaging wavelength applied on AOSH images generated at 500nm for A1 (left) and A2 (right).}
\label{lambda}
\end{figure*}
\begin{figure*}
\includegraphics[width=8.2cm]{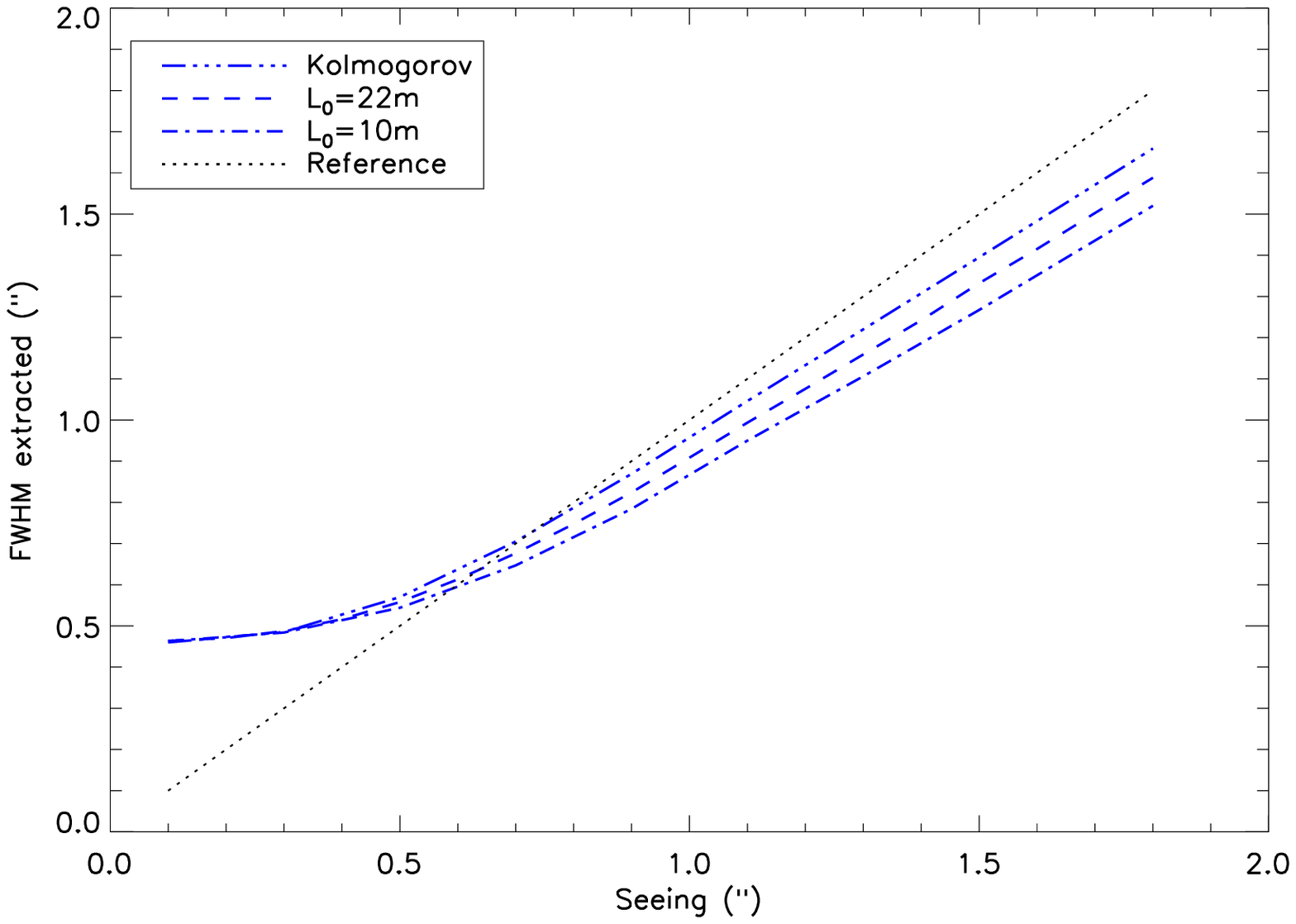}
\includegraphics[width=8.2cm]{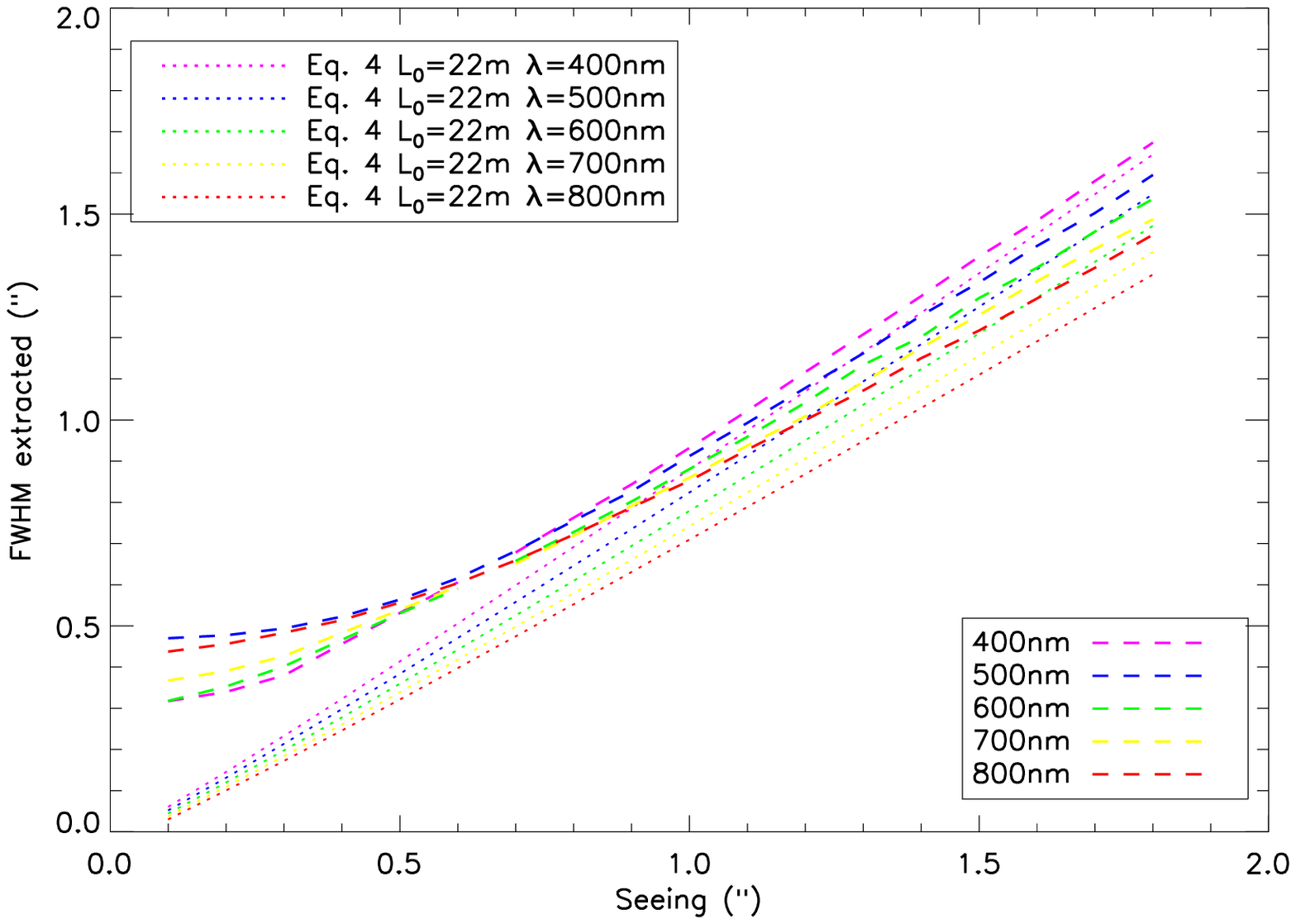}
\caption{A1 characterization: impact of the turbulence outer scale $L_0$ (left) and the AOSH imaging wavelength (right).}
\label{spot}
\end{figure*}

\subsection{Results}
In this section we provide results from the comparison of A1 and A2, while the best algorithm revealed will be further characterized against several important parameters. This calibration process is a necessary step prior to on-sky implementation of such numerical tool, and will provide insights for the re-analysis of the VLT AOSH database presented in Sect. 4.

\subsubsection{Algorithms comparison and selection}
\label{comparison}
\noindent \textit{-- Response to the seeing:} \\
\noindent Figure \ref{comp} shows the FWHM extracted with both A1 and A2 as a function of the seeing. The black dotted line corresponds to a \textit{pure} equivalence between seeing and FWHM (i.e. Eq. \ref{toko} assuming $L_0$=$\infty$), while the black dashed line corresponds to Eq. \ref{toko} with $L_0$=22m being in agreement with the statistic of the simulated atmospheric turbulence.
The general trend of both algorithms is similar although A2 constantly provides smaller FWHM estimation than A1. In addition, A1 provides a smoother response than A2 to the seeing; A2 exhibits irregularities in its behavior. \\
Two regimes are observable:
(1/) when the seeing is better than 0.6$\arcsec$ (A1) or 0.3$\arcsec$ (A2) the FWHM is higher than the seeing, which is likely due to the under-sampling of the SH (pixel scale is 0.305$\arcsec$),  (2/) when the seeing degrades further than 0.6$\arcsec$ (A1) or 0.3$\arcsec$ (A2) the FWHM is smaller and asymptotically converges towards Eq. \ref{toko} ($L_0$=22m). This last result is important as it demonstrates that although $d \ll L_{0}$, the FWHM of a long-exposure PSF obtained with very small telescope diameters (here AOSH sub-apertures of $\sim$30cm) does follow Eq. \ref{toko}. 
We note that A2 convergence to Eq. \ref{toko} ($L_0$=22m) is inaccurate for worst seeing conditions than 1.5$\arcsec$. \\

\noindent \textit{-- Sub-aperture diffraction removal:} \\
\noindent A1 and A2 differ from the way they account for the the sub-aperture diffraction. A1 is based on the deconvolution of a square sub-aperture OTF (Eq. \ref{diff1}), while A2 makes use of a quadratic subtraction of $\theta_{0}$ (Eq. \ref{diff2}). 
To compare these two approaches, we assess the sensitivity of both algorithms on the sub-aperture diffraction. In practice, A1 and A2 have been tested on the same set of AOSH images simulated at 500nm, while the input imaging wavelength used to feed the algorithms for removing the diffraction $\lambda/d$ varies from 400 to 800nm. 
Since the bandwidth center of the wavefront sensor path can vary with the guide star type, pushing this test further with such a large range of imaging wavelengths (i.e., wider amount of diffraction) is here justified. 
  
Results are presented in Fig. \ref{lambda} and show that A1 and A2 behave similarity for seeing conditions $\geq$ 1.0$\arcsec$, while for lower seeing by contrast to A1, A2 exhibits strong non reliable irregularities. 
Subtracting quadratically the diffraction FWHM $\theta_{0}$ as carried out by A2 is accurate as long as the effect of the diffraction is small relatively compared to the turbulence ($d\gg r_0$), and therefore fails at large $r_0$ (good seeing conditions).  In fact, the FWHM $\theta_{0}$ varies from 0.24$\arcsec$ to 0.49$\arcsec$ when the wavelength varies from 400 to 800nm. 
The actual AOSH PSFs are a convolution of the atmospheric blur and diffraction, and since neither of these individuals broadening factors are Gaussian, calculation of the combined FWHM as a quadratic sum of individual contributions is not accurate. \\

\noindent \textit{-- Algorithm selection:} \\
\noindent Considering all the aspects treated previously, it appears that A1 is more appropriate than A2. Therefore, in the following we will consider A1 only for further characterization for the sake of clarity until Sect. 4, where real data obtained at the VLT with A2 will be re-analyzed. The next subsections concentrate on the dependency of A1 to various parameters with the objective of allowing proper calibration of the algorithm. Nonetheless, it should be noted that A2 demonstrates identical dependency to these parameters  such that a calibration of A1 or A2 can be carried out similarly. 

\subsubsection{Turbulence outer scale}
In Fig. \ref{spot} (left) the extracted FWHM with A1 as function of the seeing is shown for several outer scale ($L_0$) values: 10, 22m, and infinite outer scale (although likely about 200 meters due to the physical finite size of the phase screens used in simulation). It is observable that A1 starts to be sensitive to the outer scale for seeing higher than 0.5$\arcsec$.
From results presented in Fig. \ref{spot} (left) it is further established (see Sect. \ref{comparison}) that AOSH sub-apertures are not small enough so that a spot FWHM can be approximated to  $\varepsilon_{0}$, i.e., the AOSH spot FWHM measurement does depend on the outer scale and therefore follows Eq. \ref{toko}. 
 
\subsubsection{Imaging wavelength}
To analyze how the AOSH imaging wavelength impacts the estimation of the seeing, we generate AOSH images for various imaging wavelength covering the visible spectrum from 400 to 800nm. Results are presented in Fig. \ref{spot} (right) and indicate that accurate knowledge of the SH imaging wavelength is critical. For all wavelengths, A1 response is constantly smooth and can therefore be efficiently calibrated. A1 response always asymptotically converges towards Eq. \ref{toko} ($L_0$=22m), where we remind the reader that $r_0$ is wavelength dependent. The shorter the wavelength, the faster the convergence to Eq. \ref{toko}. This behavior was already reported in \citet{2010A&A...516A..90M}, where we examined the validity of Eq. \ref{toko} with wavelength in the context of telescope images.
As noticed in Sect. \ref{comparison}, and generalized here for all wavelength, bellow 0.6$\arcsec$ the impact of the AOSH pixel scale is observable. The reliability of seeing estimation fails at very good seeing conditions when seeing $<$ 2$\times$ pixel scale (here 0.61$\arcsec$).

\subsubsection{Spot sampling}
To understand the asymptotical trend of the algorithm response to the seeing presented in the previous subsections, we analyze the effect of the spot sampling on the $FWHM$ measurement. 
Results are presented in Fig. \ref{sampling} (left) where it is shown that for a given seeing, the estimation gets better when the sampling improves. $FWHM$ at 0$\arcsec$/pixel sampling corresponds to theoretical values obtained with Eq. \ref{toko}. A rough estimation indicates that 5 pixels per $FWHM$ is required for accurate measurement, though it can be calibrated and corrected in case of poor sampling.

\subsubsection{Field stabilization}
The telescope field stabilization removes the low frequency Tip-Tilt components generated by, among other things, wind shacking. 
As a consequence the field stabilization also removes the slow turbulence, which may reduce the FWHM of a PSF image. 
To assess the impact of the field stabilization we compare the FWHM measurements on simulated AOSH images where the Tip-Tilt contribution (in the full telescope pupil) has been completely removed (i.e.,  perfect field stabilization with infinite bandwidth) to that of AOSH images left unmodified. This test has been carried out for several seeing conditions (0.1, 0.9, 1.8$\arcsec$) and outer scale values (10m, 22m, and infinite case). No impact at all has been revealed but at a  0.01$\arcsec$ level, which is negligible. The estimation of the seeing from the width of AOSH spots is therefore not sensitive to the field stabilization, which rises as a relevant advantage. The Tip-Tilt in the sub-aperture comes from the contribution of several low-order modes, 
which explains why the AOSH spot PSF FWHM is sensitive to the turbulence outer scale but not to field stabilization.  

\subsubsection{Atmospheric dispersion}
\begin{figure*}
\includegraphics[width=8.2cm]{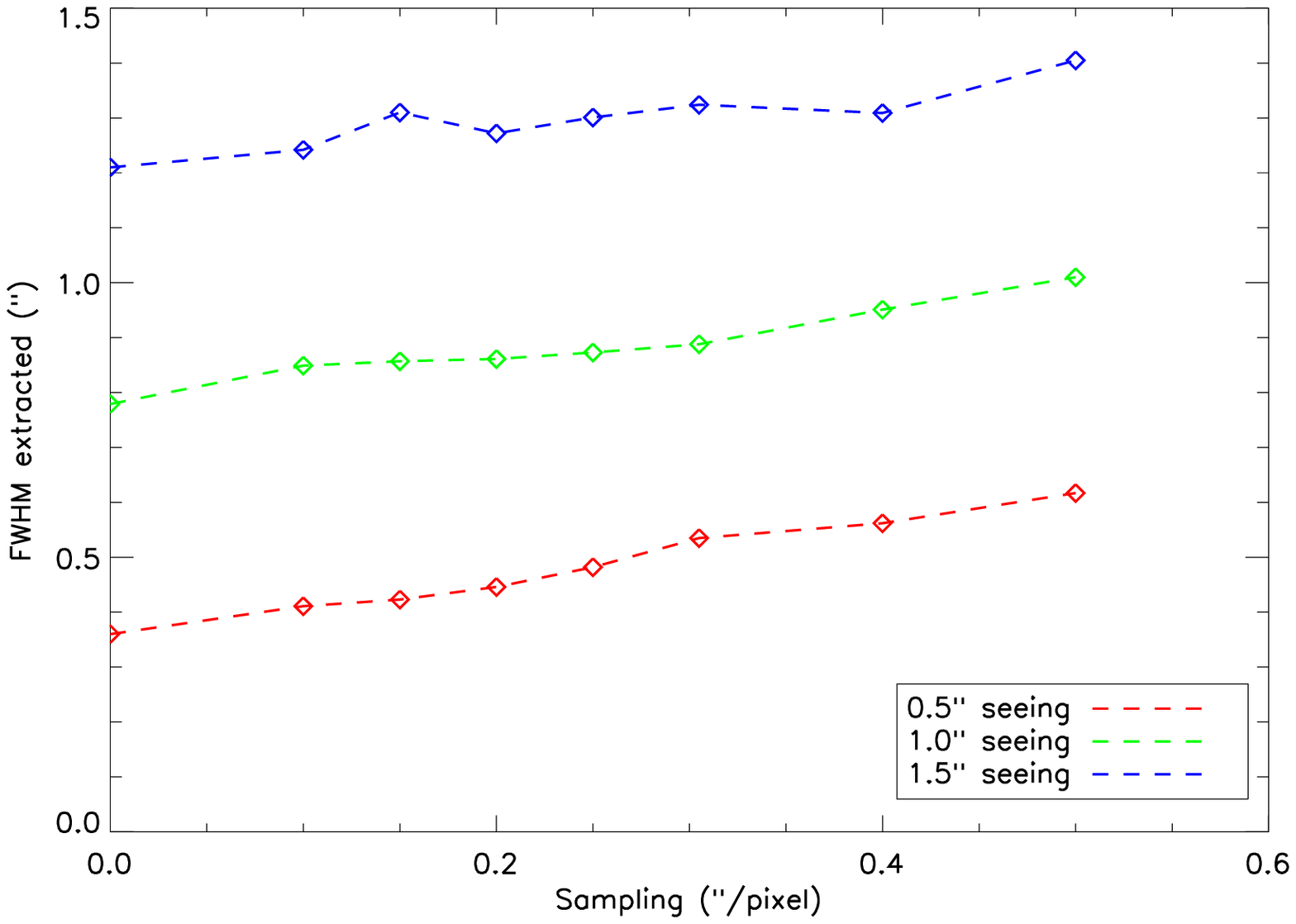}
\includegraphics[width=8.2cm]{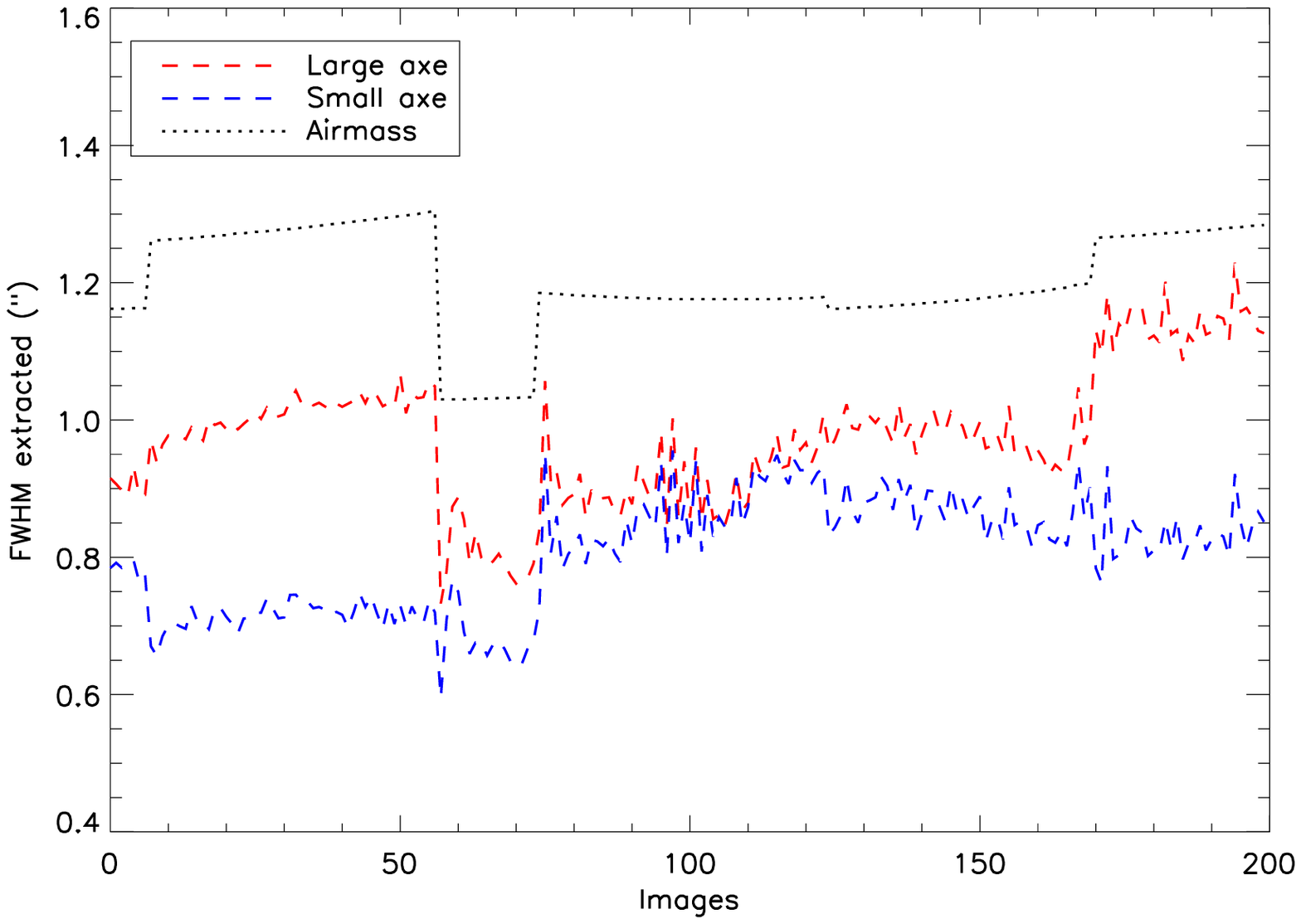}
\caption{Left: Impact of the AOSH spot sampling. Right: Impact of the uncorrected atmospheric dispersion on the seeing estimation (real data).}
\label{sampling}
\end{figure*}

Except for data taken at zenith, individual spot in the AOSH image can exhibit elongation in one direction due to uncorrected atmospheric dispersion.
A1 allows the estimation of FWHM in two orientations: the small and the large axes. 
The algorithm has been tested on 200 real AOSH images obtained at the VLT-UT3 Nasmyth focus (not equipped with an atmospheric dispersion compensator, ADC) on May 12th 2010. Results of the FWHM extracted along the small and large axis are presented in Fig. \ref{sampling} (right) where the airmass is over-plotted. Results  show that A1 does differentiate the elongated axis, and the ratio between both axis follows the evolution of the airmass. The difference between the two axis can be as large as 0.3$\arcsec$ which is substantial. The ability to distinguish small and large axis of the FWHM is therefore mandatory. We note that A2, by assuming a rotationally symmetric theoretical spot PSF profile (Eq. \ref{A2}) does not allow to differentiate these two orientations. 
\begin{figure*}
\includegraphics[width=8.2cm]{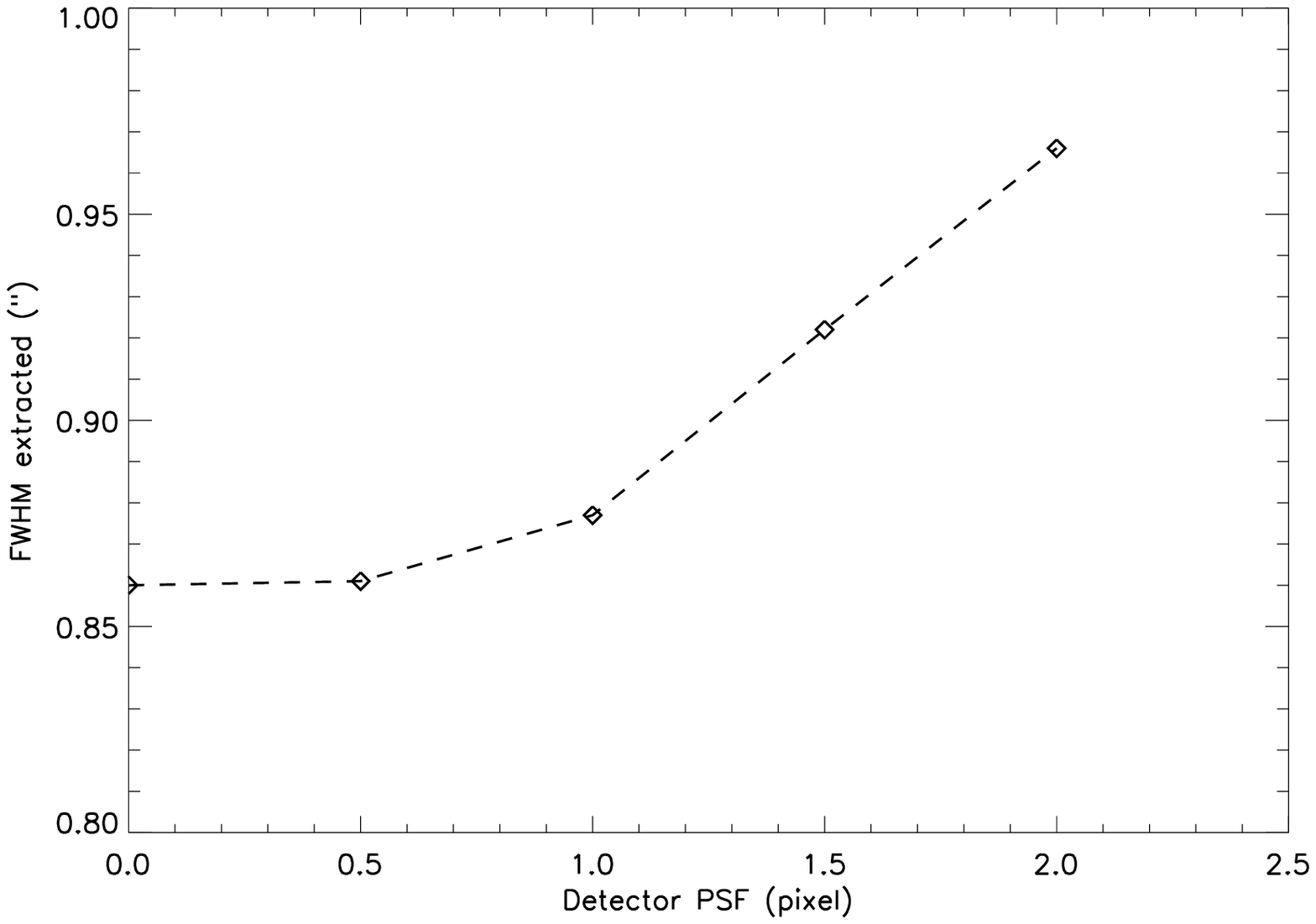}
\includegraphics[width=8.2cm]{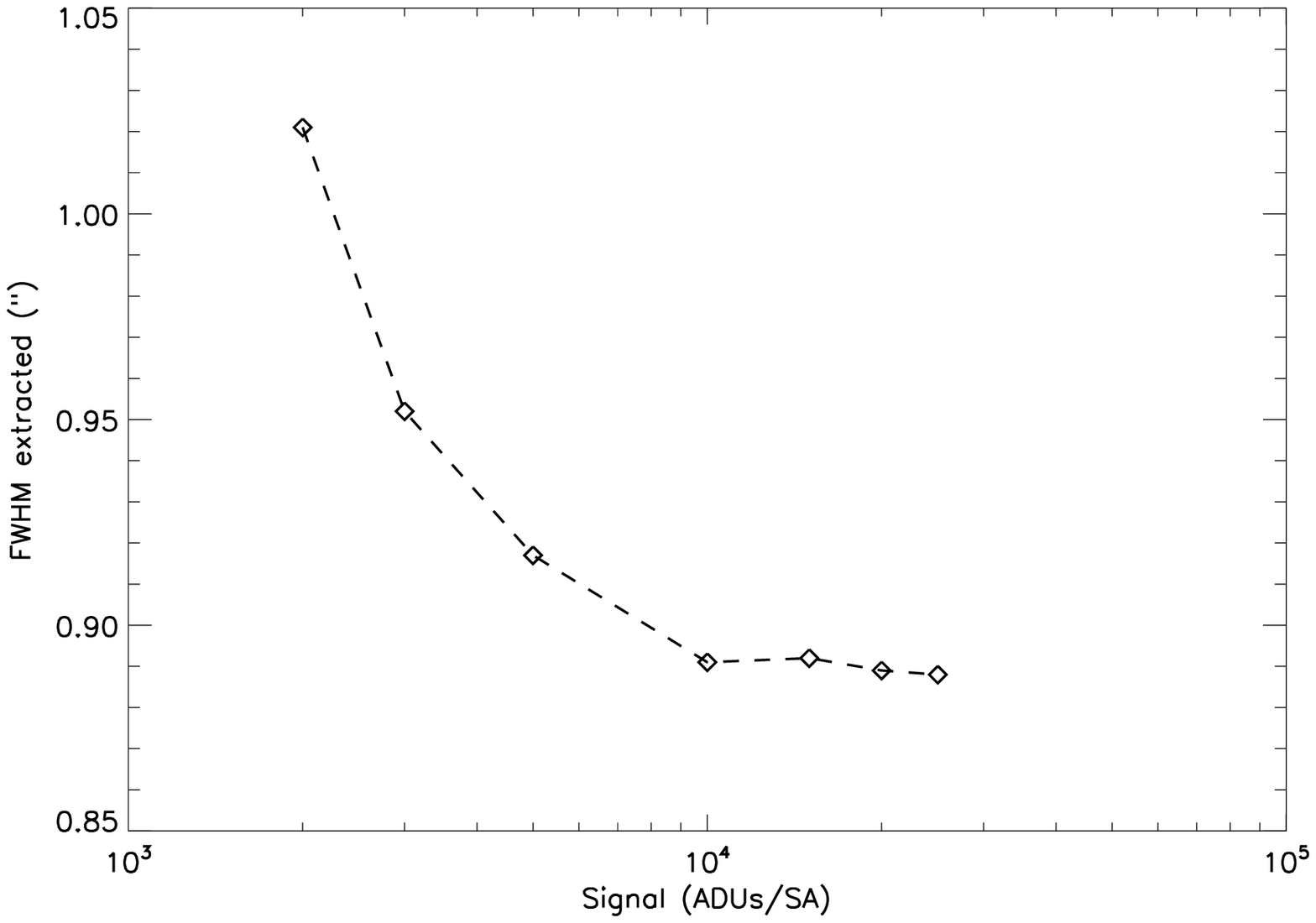}
\caption{Top: Impact of the detector PSF on the seeing estimation. Bottom: Impact of the signal-to-noise (RON=15 ADUs) on the seeing estimation.}
\label{sensitivity}
\end{figure*}

 \subsubsection{Detector PSF}
The diffusion of charges in the detector material before they are allocated to one pixel is observable as an artificial enlargement of the PSF. In most cases, the spatial response of the detector is not trivial to determine. Figure \ref{sensitivity} (left) presents the impact of the detector PSF on the measured FWHM in the case of 1.0$\arcsec$ seeing. 
We found that the detector PSF does enlarge the AOSH spot FWHM and that the effect can be significant; it starts to be substantial from 1 pixel, which is larger than what is usually encountered in scientific detectors, in particulars the ones at the VLT.
 
 \subsubsection{Signal-to-noise}
At the VLT the read-out-noise (RON) level of the AOSH is above 15 ADUs, while typical signal is about 20000 ADUs (per sub-aperture).
In figure \ref{sensitivity} (right) we present the measured FWHM in the case of 1.0$\arcsec$ seeing as a function of the signal level (in ADUs) for a RON of 15 ADUs.
It is shown that poor signal-to-noise (SNR) ratio image enlarges the FWHM which is evident for signal lower than 10000 ADUs, while for signal higher than 10000 ADUs the FWHM measurements are roughly stable. The standard signal ADU level obtained at the VLT is therefore high enough to avoid any impact on the FWHM estimation, although it can be calibrated otherwise knowing the flux.

\section{Application to the VLT AOSH database}

\subsection{Correction law and error budget}
To relate our simulations to real situation, we re-analyzed the VLT AOSH database obtained over the past year. At the VLT, FWHM estimates are recorded as diagnostic information since the commissioning of the telescope, while the A2 algorithm is used since May 2010.  
As shown in the previous sections, A2 (similarly A1) requires to be calibrated. From Fig. \ref{comp} it is straightforward to fit the data of A2, and to derive a correction law 
to transform FWHM into seeing taking into account some of the critical parameters studied in the previous sections. Accounting for the median outer scale value of Paranal  \citep[22m, ][]{2010A&A...524A..73D}, the under-sampling of the  AOSH at the VLT, and the imaging wavelength, we found that the following equation provides accurate seeing estimate from A2 FWHM measurements:
\begin{equation}
\epsilon_0 = 1.18 \times (FWHM^{1.84} - 0.15)^{1/2} 
\label{eqFIT}
\end{equation}

From the sensitivity analysis of A1 (likewise A2) to various parameters, we propose to draw here a classification of the critical parameters associated with an error budget. 
Based on the previous systematic examination of all parameters impact on the seeing estimation, we can conclude that $L_0$, $\lambda$, and the detector PSF are the major parameters to consider.
All other parameters can be excluded from the error budget, being either with negligible impact (field stabilization, SNR), with a known and constant value and thus calibrated (the spot sampling), or apart of the algorithm analysis (atmospheric dispersion). 
 The Paranal median value for the outer scale is 22m. For the error budget of this term we consider the $\pm$ one sigma deviation around the median value 22m (11m and 42m).
At the VLT, the bandwidth of the wavefront sensor path is centered around 550 nm in average. 
As it can vary with the guide star type, we consider a central wavelength of 550 nm $\pm$ 50nm.
Hence we assume here a deviation between 500 and 600nm.
For the detector PSF term, since the value is unknown at the VLT, we consider that it is confined between our reference and ideal 0 value and 1 pixel. 

\noindent We found that the error terms ($\sigma$) can be expressed as: 
\begin{equation}
\sigma_{L_0} =  0.033\times \epsilon_0,
\end{equation}
\begin{equation}
\sigma_{\lambda} = -0.009 + 0.096\times \epsilon_0 
\end{equation}
\begin{equation}
\sigma_{detector} = 0.011 \times \epsilon_0 
\end{equation}
Assuming that all terms are independent, we can therefore write the error budget of Eq. \ref{eqFIT} such as:
\begin{equation}
 \sigma^{2}_{\epsilon_0} =  \sigma^{2}_{L_0} + \sigma^{2}_{\lambda} + \sigma^{2}_{detector}
\label{eqErr}
\end{equation}
\noindent We note that for A2, by contrast with A1, an additional term should be considered: $\sigma_{AD}$, standing for the atmospheric dispersion impact. From this budget error, we can estimate that all these independent error sources affect the seeing derived from AOSH images at  $\sim$10$\%$ level. Ideally, a drastic reduction of this error level is possible through simultaneous and instantaneous measurement of $L_0$, and accurate knowledge of the guide star type.  

\subsection{Application to the VLT database}
We applied Eq. \ref{eqFIT} retroactively on the VLT AOSH database obtained at UT4. In particular, we use a set of 6500 simultaneous FWHM measurements obtained  since May 2010 with the AOSH (using A2) and the FORS2 instrument images. We remind the reader that FWHM estimation on long-exposure PSF from any scientific instrument follows Eq. \ref{toko}.  
We note that FORS2 data might be biased by several parameters, among of them the telescope field stabilization, and the accuracy of the FWHM extraction which is based on the SExtractor software \citep[Source Extractor,][]{SExtractor}. 
Briefly, SExtractor is run on the reduced FORS2 images, and stars are identified based on Sextractor parameters, while their FWHMs are measured by the program through the use of a Gaussian fit applied on the data. The FWHM is afterwards extracted from the estimated Gaussian profile.
The examination of the reliability of FORS2 data is beyond the scope of this paper, and therefore the comparison between AOSH and FORS2 data must be here, carefully considered.

\begin{figure}
\includegraphics[width=8.2cm]{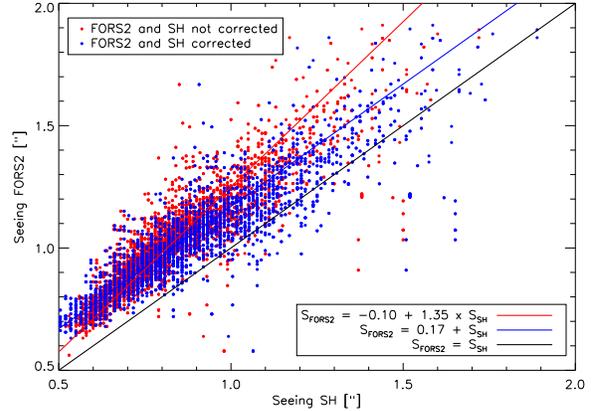}
\caption{Simultaneous seeing estimation with FORS2 and the VLT  AOSH database, while AOSH data are corrected by Eq. \ref{eqFIT} (blue dots), and left uncorrected (red dots). FORS2 data are corrected by Eq. \ref{toko} ($L_0$). Best fit of each set of data is over-plotted (full red and blue lines), while the black full line represents the reference, i.e., the ideal perfect match of FORS2 and AOSH seeing estimates.}
\label{STAT}
\end{figure}

The correction to apply to the data is carried out as follow: (1) FORS2 FWHM data are corrected by Eq. \ref{toko} assuming $L_0$=22m (Paranal median value), while A2 FWHM are left uncorrected by Eq.Ê\ref{eqFIT}, (2) FORS2 FWHM data are corrected by Eq. \ref{toko}  ($L_0$=22m) and A2 FWHM data are corrected by Eq.Ê\ref{eqFIT}. Results are shown in Fig. \ref{STAT}, where the red dots correspond to (1) and the blue dots to (2).
Figure \ref{STAT} demonstrates that FORS2 correction for $L_0$ and AOSH correction with Eq. \ref{eqFIT} bring seeing correspondence in a better agreement. This is obviously expected as both FORS2 and AOSH deliver FWHM estimate at the same location: the focus of the UT4 telescope. This result further confirm the reliability of seeing extraction from AOSH data. 

The power-laws of both set of data are provided in the legend of Fig.Ê\ref{STAT} and are over-plotted for the sake of clarity. When the AOSH data are corrected by Eq. \ref{eqFIT}, it is observable that the slope of the power-law through the data is linear and equal to unity, by contrast to the situation where AOSH data are left uncorrected. 
Only a constant and homogeneous over-estimation of FORS2 seeing remains. Such an overestimation reflects the presence of a bias, which likely finds its origin in the FWHM estimation carried out by SExtractor. For instance, long-exposure PSF profiles cannot be described by a Gaussian, and thus the calculation of the FWHM based on a Gaussian fit is not accurate. This typically induces an over-estimation of the seeing at $\sim10\%$ level.

We inform the reader that Eq. \ref{eqFIT} can be used on the VLT AOSH database, being temporally valid from May 2010 up to now. From the commissioning of the VLT  to May 2010, another correction law must be derived since another algorithm  was in use. 
In addition, in the future, A1 should replace A2 at the VLT, and thus Eq. \ref{eqFIT} will require to be re-adapted accordingly. 

\section{Conclusion}
Active Optics Shack-Hartmann (AOSH) sensor offers an advantageous possibility to estimate the seeing at the focus of a telescope: (1/) it delivers long exposure PSFs, (2/) it is not affected by  any observational bias (continuous real-time seeing estimation) by contrast to scientific instruments, (3/) we show that it is not sensitive to the telescope field stabilization.

The comparison of the two proposed algorithms gave advantage to the one proposed by \citet{Toko07} which is based on the theoretical OTF expression of the long-exposure PSF. 
Although PSF-based and OTF-based algorithms exhibit similar general behavior, we found that A1 provides a smoother response than A2 to the seeing and better account for the sub-aperture diffraction, and atmospheric dispersion.

We show that the estimation of the seeing from AOSH images is sensitive to several parameters but can be calibrated.
We establish that even considering the small size of the AOSH sub-apertures ($\sim$30cm), it does depend on the turbulence outer scale $L_{0}$ and therefore follows Eq. \ref{toko}. We show that AOSH can be used to build statistics using median value of $L_{0}$, in a real-time fashion but relying on median $L_{0}$ value, or requiring instantaneous measurement of $L_{0}$. We note that in this context several independent campaigns have converged to the median value of $L_{0}$ of 22 m at Paranal \citep[e.g.,][]{2010A&A...524A..73D}. 

The VLT AOSH database available since the commissioning of the telescope does not provide seeing information but spot FWHMs. In this context, Eq. \ref{eqFIT} is required to derive the seeing information. We remind the reader, that Eq. \ref{eqFIT} is temporally valid until May 2010.
With these considerations in mind, the re-analysis of the past years of the VLT AOSH database demonstrates a seeing correspondence in better agreement between AOSH and the FORS2 imager. This result further confirms the reliability of seeing extraction from AOSH images. 

Finally, the qualification and calibration of the algorithm A1 are nearly completed and clear the path to its operational implementation at the VLT for three months of test period starting in fall 2011. Finally, we note that more emphasis is given to use closed-loop real-time AO instruments data in the near future to get the estimate of the seeing at the critical location of the telescope focus. 

\section*{Acknowledgments}
The activity outlined in the paper was supported by the European Commission, Seventh Framework Programme (FP7), Capacities Specific Programme, Research Infrastructures; specifically the FP7, Preparing for the construction of the European Extremely Large Telescope Grant Agreement, Contract number INFRA-2007-2.2.1.28.
We acknowledge Philippe Duhoux from ESO for his help with the VLT control software.

\end{document}